\documentclass{llncs}

\usepackage[dvips]{epsfig}

\begin{document}

\def\bottomfraction{.5}

\title{Density fluctuations and phase separation in a traffic flow model}

\author{S.~L\"ubeck \and M.~Schreckenberg \and {K.\,D.}~Usadel}

%\author{S.~L\"ubeck\inst{1} \and M.~Schreckenberg\inst{1} 
%\and {K.\,D.}~Usadel\inst{1}}

\institute{Theoretische Physik, Gerhard-Mercator-Universit\"at, 
47048 Duisburg, Deutschland
%\and
%Universit\'{e} de Paris-Sud,
%Laboratoire d'Analyse Num\'{e}rique, B\^{a}timent 425,\\
%F-91405 Orsay Cedex, France
}

\maketitle

\markboth{\rm \tiny{
in {\it Traffic and Granular Flow 97},
edited by D.~E.~Wolf and M.~Schreckenberg,
Springer, Singapore (1998)}}
{\rm \tiny{
in {\it Traffic and Granular Flow 97},
edited by D.~E.~Wolf and M.~Schreckenberg,
Springer, Singapore (1998)}}
\thispagestyle{myheadings}
\pagestyle{myheadings}

\begin{abstract}
Within the Nagel-Schreckenberg traffic flow model we consider 
the transition from the free flow regime to the 
jammed regime.
We introduce a method of analyzing the data which is based 
on the local density distribution.
This analyzes allows us to determine the phase diagram
and to examine the separation
of the system into a coexisting free flow phase
and a jammed phase above the transition.
The investigation of the steady state structure factor
yields that the decomposition in this phase coexistence regime
is driven by density fluctuations, provided they exceed
a critical wavelength.
\end{abstract}

\section{Introduction}

Over the past few years much attention has been devoted to 
the study of traffic flow. 
Since the seminal work of Lighthill and Whitham 
in the middle of the 50's \cite{LIGHT_1} many attempts have been
made to construct more and more sophisticated models which incorporate
various phenomena occurring in real 
traffic (for an overview see \cite{WOLF}).
Recently, a new class of models, based on the idea of cellular
automata, has been proven to describe traffic dynamics
in a very efficient way \cite{NASCH_1}.
Especially the transition from free flow to jammed
traffic with increasing car density could be investigated
very accurately.
Nevertheless, besides various indications \cite{CSANYI_1}, 
no unique description for a dynamical transition has 
been found (see for instance \cite{SASVARI_1,LUEB_6} and
references therein).
%Furthermore, no satisfying order parameter could be defined
%so far.
In this article we consider a method of analysis
which allows us to identify the different phases of the system
and to describe the phase transition, i.e., considering the 
fluctuations which drive the transition,
and determining the phase diagram.

We consider a one-dimensional cellular automaton of
linear size $L$ and $N$ particles.
Each particle is associated the integer values 
$v_i\in\{0,1,2,...,v_{\rm max}\}$ and $d_i\in\{0,1,2,3,...\}$,
representing the velocity and the distance to the next
forward particle \cite{NASCH_1}.
For each particle, the following 
update steps representing the acceleration, the slowing down, the noise,
and the motion of the particles are done in parallel:
(1) if $v_i < d_i$ then $v_i \to \mbox{Min}\{v_i+1, v_{\rm max}\}$,
(2) if $v_i > d_i$ then $v_i \to d_i$,
(3) with probability $P\/$ $v_i \to \mbox{Max}\{v_i-1, 0\}$, 
and
(4) $r_i \to r_i+v_i$, 
where $r_i$ denotes the position of the $i$-th particle.

\section{Simulation and Results}

\begin{figure}[b]
 \begin{center}
 \epsfxsize=8.6cm
 \epsfysize=7.0cm
% \epsffile{/nfs/nujunkum/sven/traffic/paper_1/space_time_dia.eps} 
 \epsffile{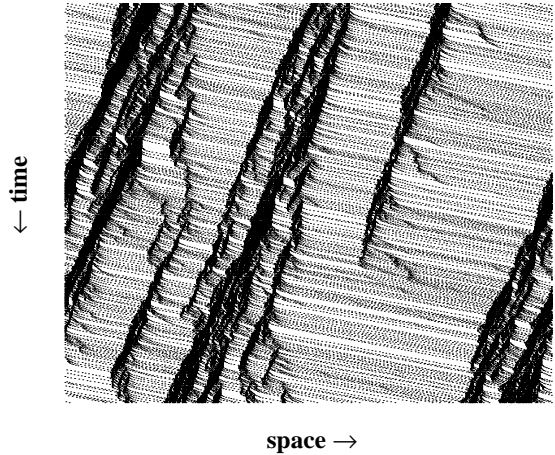} 
 \end{center}
 \vspace{-0.8cm}
 \caption{Space-time plot for $v_{\rm max}=5$, $P=0.5$, and $\rho_g>\rho_c$.
          Note the separation of the system in high and low density
	  regions.}
 \label{space_time_dia} 
\end{figure}

Figure~\ref{space_time_dia} shows a space-time plot of the system.
Each dot corresponds to a particle at a given time step.
The global density $\rho_g=N/L$ exceeds the critical density and 
jams occur.
Traffic jams are characterized by a high local density of the
particles and by a backward movement of shock waves \cite{LIGHT_1}.
One can see from Fig.~\ref{space_time_dia} that in the jammed regime 
the system is inhomogeneous, 
i.e., traffic jams with a high local density and free flow regions with
a low local density coexist.
In order to investigate this transition one has to take this
inhomogeneity into account.

Traditionally one determines the so-called fundamental diagram, 
i.e.,~the diagram of the flow vs the density.
The global flow is given by, 
%\begin{equation}
$\Phi \; = \; \rho_g \, \langle v \rangle$,
%\label{eq:flow_density}
%\end{equation}
where $\langle v \rangle$ denotes the averaged velocity of the 
particles.
This non-local measurements are not sensitive to the inhomogeneous
character of the system, i.e.,~the information
about the two different coexisting phases is lost.
In the following we consider
a method of analysis which is based on the measurement of 
the local density distribution $p(\rho)$~\cite{LUEB_6}.
The local density $\rho$ is measured on a section of the 
system of size $\delta$ according to
\begin{equation}
\rho \; = \; \frac{1}{\rho_g \delta} \, \sum_{i=1}^{N} \,
\theta(\delta -r_i).  
\label{eq:density_dist}
\end{equation}
%Of course we have checked that the main results are not
%affected by the value of $\delta$, provided that delta
%is significantly smaller than the system size $L$ in 
%order to measure the local properties.
%In order to reflect the behavior of the low density
%regime $\delta$ should be significantly larger
%than a certain length scale $\lambda_0$ which corresponds 
%to the characteristic length scale of the density 
%fluctuations in the free flow phase (see below).
%For any parameter set $\{v_{\rm max},P\}$ 
%the local density $\rho$ fluctuates around the value of the
%global density $\rho_g$ and the probability distribution 
%of the local density $p(\rho)$ contains all informations
%needed to describe the transition.

\begin{figure}[p]
 \begin{center}
 \epsfxsize=8.0cm
 \epsfysize=8.0cm
 \epsffile{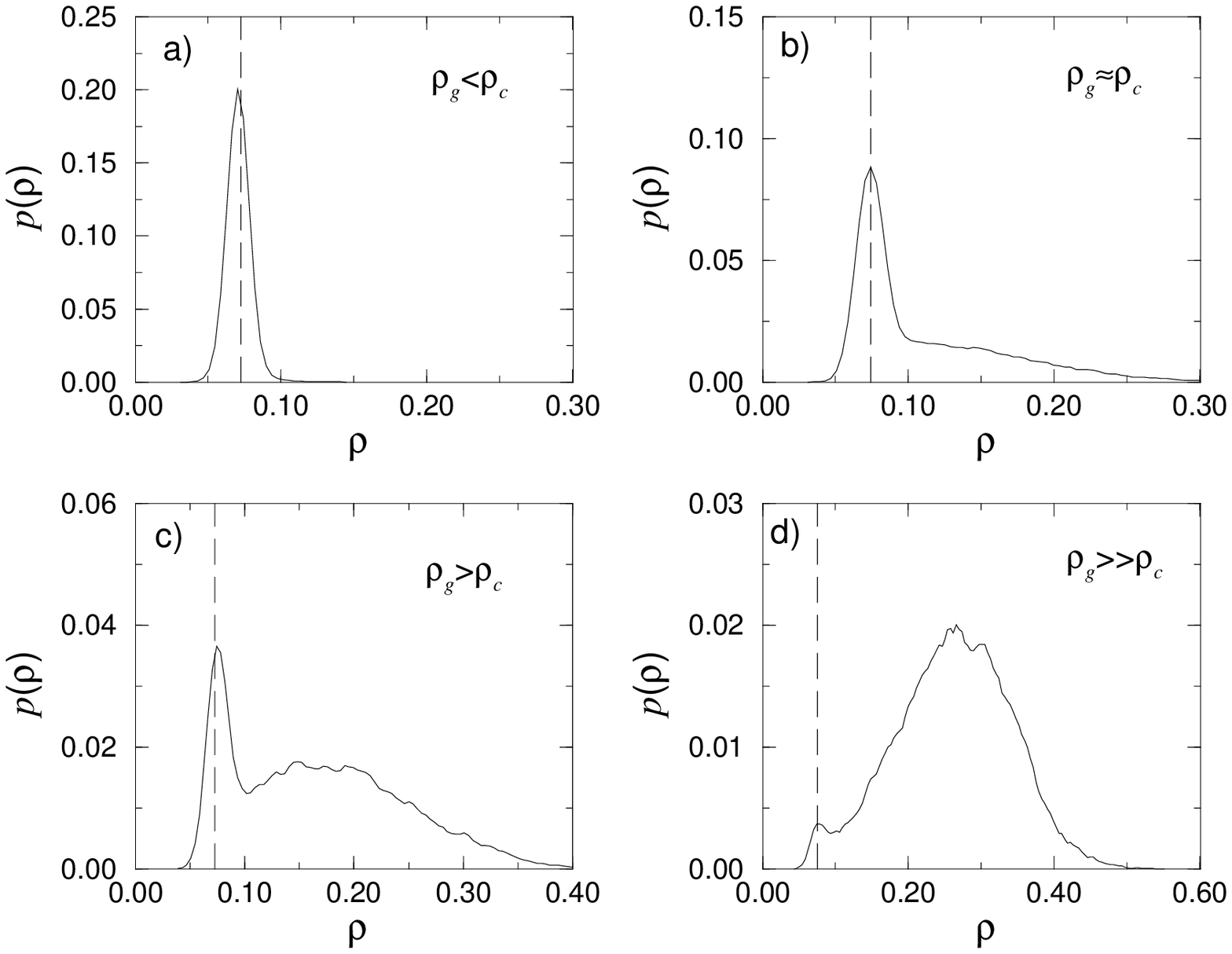} 
 \end{center}
 \vspace{-0.9cm}
 \caption{The local density distribution $p(\rho)$ for various
	  values of the global density, $v_{\rm max}=5$, $P=0.5$
	  and $\delta=256$. 
	  The dashed line corresponds to the characteristic density 
	  of the free flow phase.}
 \label{density_dist} 
 \vspace{0.5cm}
%\end{figure}
%\begin{figure}[b]
 \begin{center}
 \epsfxsize=7.4cm
 \epsfysize=7.4cm
% \epsffile{/nfs/nujunkum/sven/traffic/figures/b_8192_0128_2.ps} 
 \epsffile{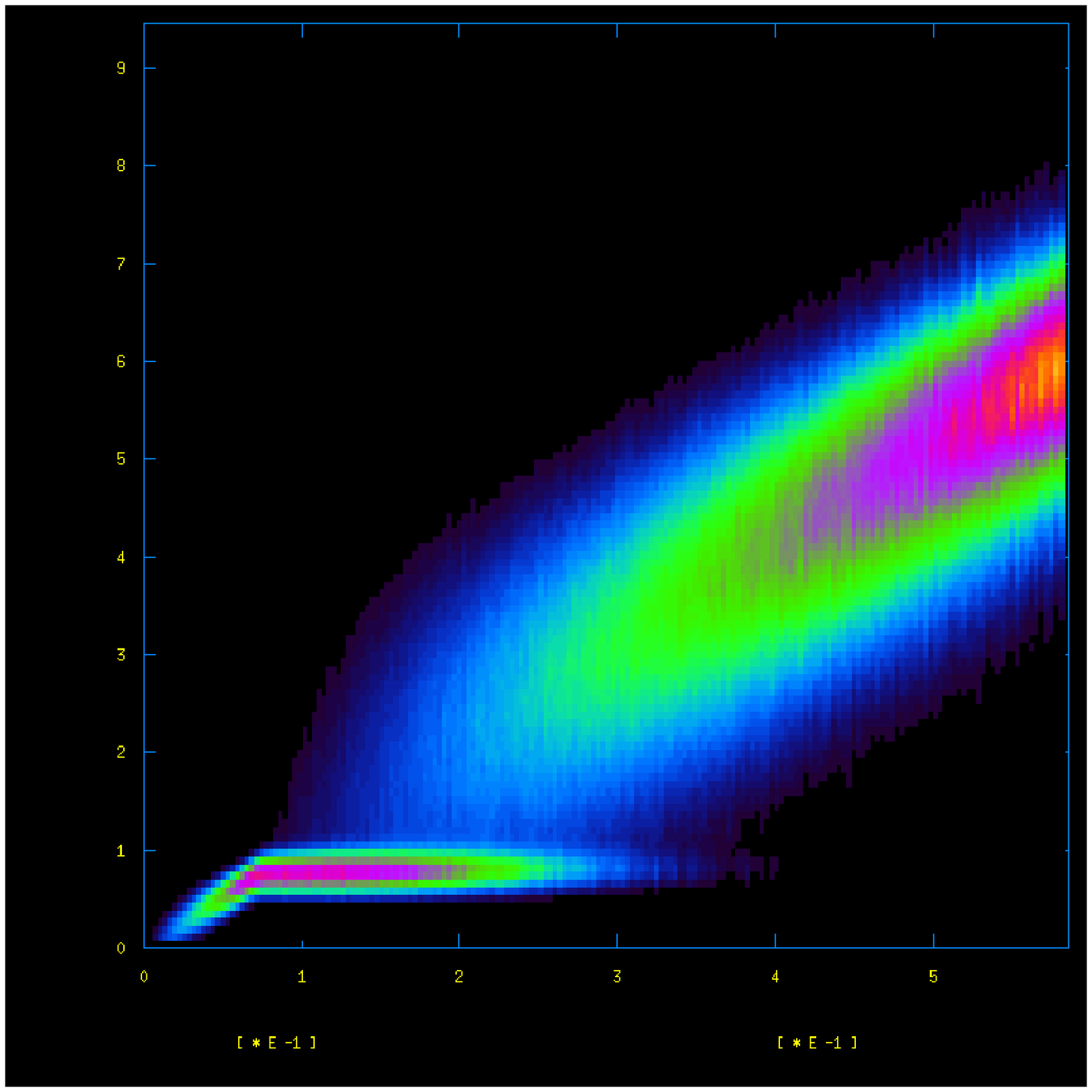} 
 \end{center}
 \vspace{-0.0cm}
 \caption{The local density distribution $p(\rho_g,\rho)$ as a 
	  function of the global density (horizontal axis) and 
	  local density (vertical axis), respectively.
	  The colors correspond to the values of the probability 
	  $p(\rho_g,\rho)$, increasing from black to red.}
 \label{density_dist_2} 
\end{figure}

The local density distribution $p(\rho)$ is plotted 
for various values of the global 
density $\rho_g$ in Fig.~\ref{density_dist}.
In the case of small values of $\rho_g$, see Fig.~\ref{density_dist}a, 
the particles can
be considered as independent (see below) and the 
local density distribution is simply Gaussian with
the mean values $\rho_g$ and a width which scales 
with $\sqrt{\delta}$. 
Increasing the global density, jams occur
and the distribution displays two different 
peaks (Fig.~\ref{density_dist}c).
The first peak corresponds to the density of free particles and
in the phase coexistence regime the position of this peak
does not depend on the global density (see the dashed lines
in Fig.~\ref{density_dist}).
The second peak is located at larger densities and characterizes
the jammed phase.
With increasing density the second peak occurs in the vicinity
of the critical density $\rho_c$ (Fig.~\ref{density_dist}b)
and grows further (Fig.~\ref{density_dist}c) until it dominates
the distribution in the sense that the first peak 
disappears (Fig.~\ref{density_dist}d).
The two peak structure of the local density distribution clearly 
reflects the coexistence of the free flow and 
jammed phase above the critical value $\rho_c$.
In Fig.~\ref{density_dist_2} we present the 
probability distribution as function of the
global and local density.
Above a certain value of the global density $\rho_g$  
the two peak structure occurs.
The behavior of the first peak yields a criterion to 
determine the transition point~\cite{LUEB_6} and one gets
$\rho_c=0.0695\pm 0.0007$ for $P=0.5$ and $v_{\rm max}$,
respectively.

%In Fig.~\ref{density_dist} we plot the position of the maximum
%of the local density distribution $\rho(p_{\rm max})$ as a
%function of the global density $\rho_g$.
%One clearly sees the transition point $\rho_c$ where
%the position of the maximum becomes independent of the
%global density.
%The inset of Fig.~\ref{density_dist} shows that the
%determination of the transition point does not
%depend on the special value of $\delta$.
%Only the point where the second peak exceeds the first peak
%depends on the measurement parameter $\delta$.
%With increasing $\delta$ this point tends to smaller values 
%of $\rho_g$ because with increasing $\delta$ the measurement
%starts to average over the two different phases.
%From these measurements we conclude that the 
%phase transition of the Nagel-Schreckenberg model
%is a transition from a homogeneous regime (free flow phase)
%to an inhomogeneous regime which is characterized by a 
%coexistence of two phases (free flow traffic and jammed traffic).

\begin{figure}[t]
 \begin{center}
 \epsfxsize=8.6cm
 \epsfysize=7.5cm
 \epsffile{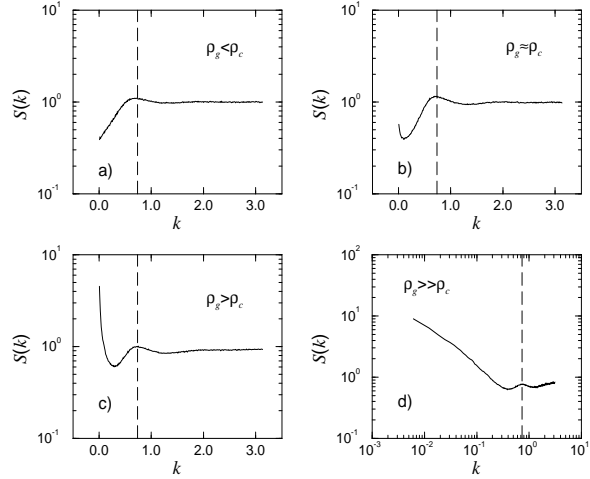} 
 \vspace{-0.8cm}
 \caption{The structure factor $S(k)$ for $P=0.5$, $v_{\mbox{max}}=5$
	  and for various values of the
	  density $\rho$. The dashed line marks the characteristic
	  wavelength $\lambda_0$ of the free flow phase.}
 \label{struc_5} 
 \end{center}
\end{figure}

In order to describe the spatial decomposition of the
coexisting phases we measured the steady state structure factor \cite{SCHMITT_1}
\begin{equation}
S(k) \; = \; \frac{1}{L} \left \langle \left | \, \sum_{r=1}^{L} \,
\eta(r) \, e^{i k r} \right|^2 \right \rangle,
\label{eq:structure_factor}
\end{equation}
where $\eta(r)=1$ if the lattice site $r$ is occupied and
$\eta(r)=0$ otherwise.
In Fig.~\ref{struc_5} we plot the structure factor $S(k)$
for the same values of the global density as in Fig.~\ref{density_dist}, 
i.e., below, in the vicinity, above and far away of the transition
point.
It is remarkable that $S(k)$ exhibits a maximum for all considered
values of the global density at $k_0 \approx 0.72$ (dashed lines in
Fig.~\ref{struc_5}). 
This value correspondence to the characteristic wave length 
$\lambda_0=\frac{2 \pi}{k_0}$ of the density fluctuations 
in the free flow phase.
The steady state structure factor is related to the 
Fourier transform of the real space density-density
correlation function.
The wave length $\lambda_0$ corresponds to a maximum of the
correlation function, i.e., $\lambda_0$ describes the 
most likely distance of two particles in the free flow phase.
For low densities the structure factor is almost independent
of the density and displays a minimum
for small $k$ values indicating the lack of long-range correlations.
Crossing the transition point the smallest mode 
$S(k=\frac{2 \pi}{L})$ increases quickly. 
This suggests that the jammed phase is characterized by 
long-range correlations which decay in the limit $\rho_g \gg \rho_c$ 
algebraically as one can see from the log-log plot 
in Fig.~\ref{struc_5}d.

\begin{figure}[t]
 \begin{center}
 \epsfxsize=8.6cm
 \epsfysize=7.0cm
 \epsffile{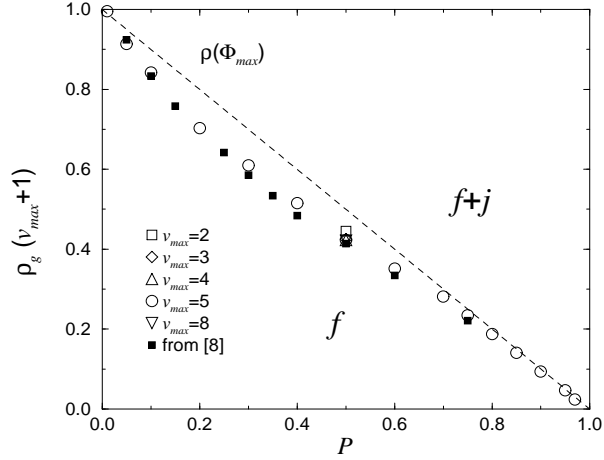} 
 \end{center}
 \vspace{-1.0cm}
 \caption{The phase diagram of the Nagel-Schreckenberg model.
	  Note that in the non-deterministic region $0<P<1$ 
	  the density of the maximum flow exceeds 
	  the density of the transition point.}
 \label{phase_dia}
\end{figure}

Up to now we only considered the case $P=0.5$.
The phase diagram in Fig.~\ref{phase_dia} shows the $P$ dependence 
of the transition density $\rho_c$.
{\it f} denotes the free flow phase and
{\it f+j} corresponds to the coexistence region
 where the system separates in the
free flow and jammed phase.
The dashed line displays the $P$ dependence of the maximum
flow obtained from an analysis of the fundamental
diagram \cite{EISEN_1}.
The critical densities $\rho_c$, where the phase transition 
takes place, are lower than 
the density values of the maximum flow.
Measurements of the relaxation time, which is expected to
diverge at a transition point \cite{CSANYI_1},
confirm this result \cite{EISEN_2} (see Fig.~\ref{phase_dia}).
But one has to mention that the determination of the critical
density via relaxation times leads in the coexistence regime
{\it f+j} to unphysical results, in the sense
that the relaxation time becomes negative \cite{EISEN_1,EISEN_2}.

\section{Conclusions}

In conclusion we have studied numerically the Nagel-Schreckenberg 
traffic flow model using a local density analysis.
Crossing the critical line of the system
a phase transition takes place from a 
homogeneous regime (free flow phase)
to an inhomogeneous regime which is characterized by a 
coexistence of two phases (free flow traffic and jammed traffic).
The decomposition in the phase coexistence regime
is driven by density fluctuations, provided they exceed
a critical wavelength $\lambda_c$.

\end{document}